\documentclass[%
 aip,
 jmp,%
 amsmath,amssymb,
 reprint,%
 floatfix,
]{revtex4-1}

\usepackage{graphicx}
\usepackage{dcolumn}
\usepackage{bm}

\usepackage{braket}            
\usepackage{color}
\usepackage[dvipsnames]{xcolor}
\usepackage{placeins}
\usepackage{float}
\usepackage{multirow}

\graphicspath{{images/}}

\newcommand{\dd}{\mathrm{d}}
\newcommand{\kB}{k_\mathrm{B}}
\newcommand{\kT}{\kB T}

\newcommand{\erf}{\text{Erf}}

\newcommand{\QA}{Q_\mathrm{A}}

\newcommand{\be}{\begin{equation}}
\newcommand{\ee}{\end{equation}}
\newcommand{\bea}{\begin{eqnarray}}
\newcommand{\eea}{\end{eqnarray}}

\begin{document}

\title{Transition path time distributions}

\author{M. Laleman}
\affiliation{KU Leuven, Institute for Theoretical Physics, Celestijnenlaan
200D, 3001 Leuven, Belgium}

\author{E. Carlon}
\affiliation{KU Leuven, Institute for Theoretical Physics, Celestijnenlaan
200D, 3001 Leuven, Belgium}

\author{H. Orland}
\affiliation{Institut de Physique Th\'eorique, CEA, CNRS, UMR3681, F-91191 Gif-sur-Yvette, France}
\affiliation{
Beijing Computational Science Research Center, No.10 East Xibeiwang Road, Beijing 100193, China}
\email{henri.orland@cea.fr}

\date{\today}

\begin{abstract}
Biomolecular folding, at least in simple systems, can be described as
a two state transition in a free energy landscape with two deep wells
separated by a high barrier. Transition paths are the short part of the
trajectories that cross the barrier. Average transition path times and,
recently, their full probability distribution have been measured for
several biomolecular systems, e.g. in the folding of nucleic acids
or proteins.  Motivated by these experiments, we have calculated
the full transition path time distribution for a single stochastic
particle crossing a parabolic barrier, focusing on the {\sl underdamped}
regime. Our analysis thus includes inertial terms, which were neglected in
previous studies. These terms influence the short time scale dynamics of
a stochastic system, and can be of experimental relevance in view of the
short duration of transition paths.  We derive the full transition path
time distribution in the underdamped case and discuss the similarities
and differences with the high friction (overdamped) limit.
\end{abstract}

\maketitle

\section{Introduction}

\begin{figure}[b]
\includegraphics[width = 0.4 \textwidth]{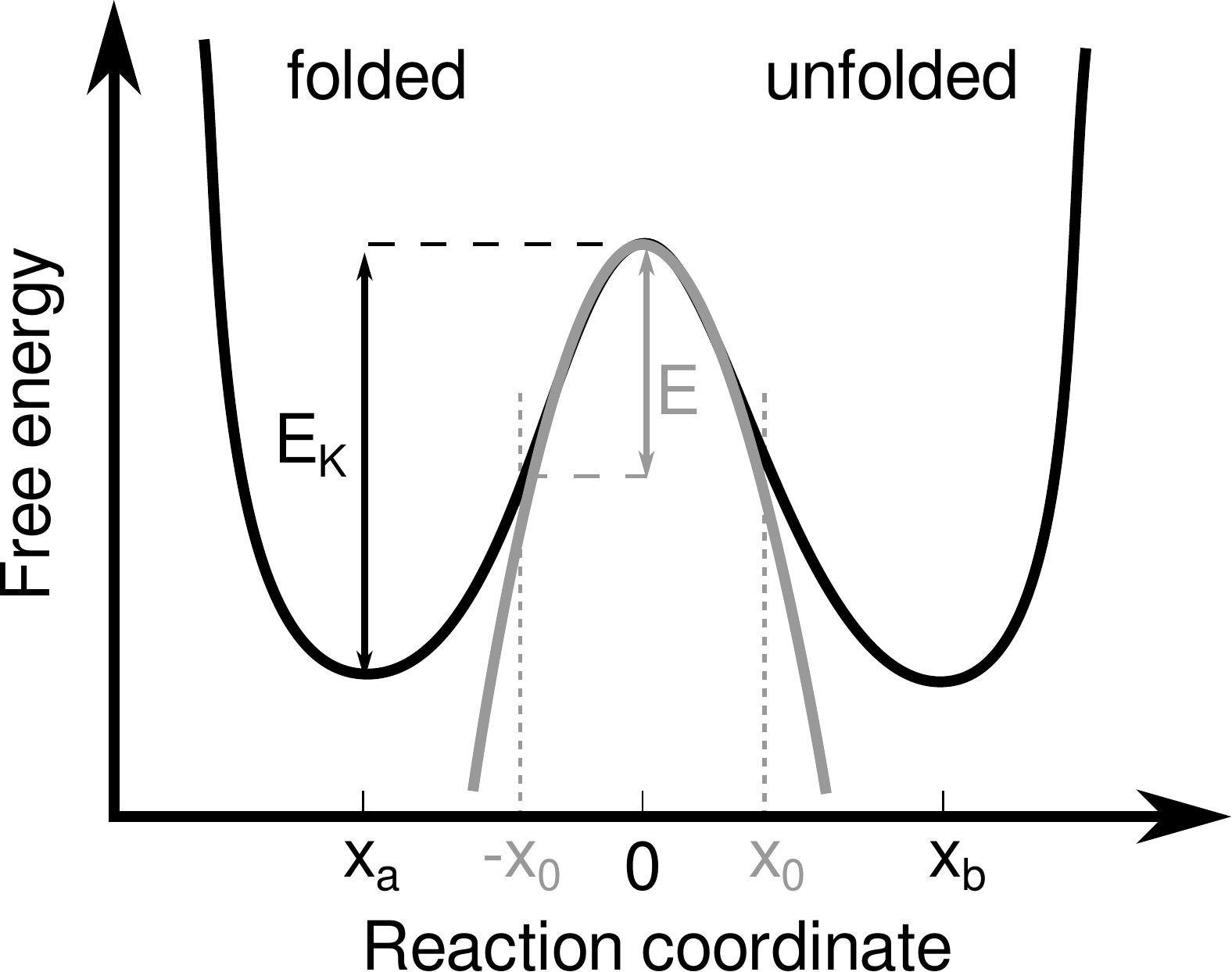}
\caption{The complex motion of a system is often mapped onto the motion of a 
single \textit{reaction coordinate}. Two-state systems are then imagined as the 
movement of this reaction coordinate in a one-dimensional free energy 
landscape, with two minima separated by an energy barrier $E$. To calculate the 
transition path time distribution, the top of the barrier is modelled as a 
parabolic barrier.}
\label{fig:landscape}
\end{figure}

Conformational changes of macromolecules between two different states
are usually described by a transition in a double well potential
landscape~\cite{hang90} (see Fig.~\ref{fig:landscape}). In this
description one assumes that the complex collective behavior of
the system can be mapped onto a single reaction coordinate. It is
also typically assumed, as done in this work, that the reaction
coordinate follows a stochastic memoryless Markovian dynamics
(for a discussion of memory effects in molecular folding, see
e.g.~\cite{walt12,saka17,sati17,vand17}).

If the barrier is high compared to the characteristic thermal energy
$k_BT$, the molecule spends most of the time close to the minima, while
the transition region around the barrier top is rarely visited. Transition
paths are the part of the trajectories spent crossing the potential
barrier connecting the two wells.  Due to the very short duration
of these trajectories, their experimental analysis has been a big
challenge for a long time, but measurements of transition path times
in nucleic acids and protein folding have been successfully performed
in the past decade~\cite{chun09,neup12,true15,neup17}. Transition
paths have also attracted the attention of theorists
and their properties were discussed in several
papers~\cite{humm04,bere05,dudk06,zhan07,sega07,chau10,orla11,fred14,kim15,maka15,dald16,bere17}.

So far most of the theoretical studies focused on models of single
particles crossing a one dimensional potential barrier in the overdamped
regime in which friction forces dominate over inertia. In ordinary
macromolecular dynamics, the high friction (overdamped) regime applies at
typical experimentally sampled timescales. However, in view of the very
short duration of transition paths, it is interesting to explore the
underdamped regime in which inertial effects are taken into account. The
aim of this paper is to discuss this regime.

We present analytical calculations of the distribution of transition
path times (TPT's) based on the Langevin equation with Gaussian white
noise for a particle of mass $m$, and friction coefficient $\gamma$,
which crosses a parabolic barrier $V(x)=-Kx^2/2$. In particular, we
emphasize the differences with the TPT distribution in the overdamped
limit, which has been already studied in the past~\cite{zhan07}.
The manuscript is organized as follows: Section~\ref{sec:Kramers} briefly
recalls the setup for the analysis of overdamped particle dynamics
and gives the underdamped Kramers' time.  Section~\ref{sec:TPT_gen}
discusses the general setup of the problem and links the TPT probability
distribution to an absorption probability. Section~\ref{sec:barrier}
shows the details of the calculation of the TPT distribution for
a parabolic barrier and discusses several regimes of the obtained
expression. Section~\ref{sec:av_TPT} discusses the behavior of the
average TPT, while Section~\ref{sec:numerics} shows a comparison between
the analytical results and numerical simulations.

\section{Review of underdamped dynamics}
\label{sec:Kramers}

To analyze the underdamped dynamics of a particle of mass $m$, subject
to an external force $F(x)$ and to a frictional force $-\gamma v$
we will make use of the Langevin equation~\cite{risk84}
\begin{equation}
\label{eq:Lang}
m\ddot{x} = F(x) -\gamma \dot{x}  + \eta(t),
\end{equation}
where the dot indicates the time derivative and $\eta(t)$ is a Gaussian
random force with properties
\begin{equation}
\label{eq:gaussian_white_noise}
\begin{cases}
\langle \eta(t) \rangle  = 0, \\
\langle \eta(t) \eta(t^\prime) \rangle  = 2 \kT \gamma \delta(t - t').
\end{cases}
\end{equation}
Here $\langle . \rangle$ denotes the average over different stochastic
realizations.

Before discussing TPT it is convenient to recall some known results about
Kramers' time, which is the characteristic time for the particle to jump
from one well to the other in a double well potential $V(x)$ as that
shown in Fig.~\ref{fig:landscape}. This is obtained from the solution
of Eq.~\eqref{eq:Lang}, or of the associated Fokker-Planck equation.
In the underdamped case the time for jumping from well $a$ to well $b$
is given by~\cite{hang90}
\begin{equation}
\tau_{a\to b} = 2\pi \sqrt{\frac{K}{K_a}} \,
\frac{2m \, e^{\displaystyle \beta E_K}}{\sqrt{\gamma^2 + 4Km} - \gamma},
\label{eq:Kramers2}
\end{equation} 
where $E_K$ is the barrier height and $\beta=1/k_BT$. The Kramers'
time depends also on the curvature of the potential energy $V(x)$ at
the bottom ($K_a$) and at the top ($K$) of the barrier, obtained from
the expansions $V(x) \approx K_a (x-x_a)^2 /2$ and $V(x) \approx -K
x^2 /2$. The overdamped regime can be obtained by letting $m \to 0$
in~\eqref{eq:Kramers2}, which yields
\begin{equation}
\lim_{m \to 0} \tau_{a\to b} = 
\frac{2\pi \gamma}{\sqrt{K_a K}}\, e^{\displaystyle \beta E_K}
\equiv \tau^{(o)}_{a\to b} 
\label{eq:Kramers}
\end{equation}
Note that the underdamped dynamics is always slower than the overdamped
dynamics as $\tau > \tau^{(o)}$, i.e. inertia has the effect of slowing
down the dynamics~\cite{hang90}. Characteristic of the Kramers' time is
the exponential dependence on the barrier height.

\section{Transition path times distributions}
\label{sec:TPT_gen}

In this Section we present the general framework for the calculation
of $p_\text{TP}(t)$, the probability distribution that a transition
path has a duration $t$. To define such paths one needs to fix the
values of the reaction coordinates at the two sides of the barrier,
as illustrated in Fig.~\ref{fig:landscape}.  There is some freedom
in the choice of the original and final point of transition paths,
and the TPT distribution might depend on it.  For convenience we place
the maximum at $x=0$ and we consider paths from $-x_0<0$ to $x_0>0$.
Transition paths are trajectories connecting the initial to the final
point without entering the regions $x<-x_0$ and $x>x_0$.

The key quantity is the propagator $\mathcal{P}(x, v, t|x_0, v_0, 0)$,
which is the probability that a particle with initial position and
velocity $x_0$ and $v_0$ is at $x$ and has velocity $v$ at a later time
$t > 0$. The calculation of TPT distributions requires thus absorbing
boundary conditions at $-x_0$ and $x_0$. Consider a path originating
at a point $x'<x_0$ and with velocity $v'$; in the underdamped case an
absorbing region at $x>x_0$ means that~\cite{burs81,risk84}
\begin{equation}
\mathcal{P}(x_0, v, t|x', v', 0) = 0 \qquad \text{for}\quad v<0
\label{P2}
\end{equation}
which corresponds to setting to zero the current of particles leaving
the wall towards the domain. Note that this is more subtle than the
overdamped case for which more simply $\mathcal{P}(x_0, t|x', 0) = 0$.
The condition \eqref{P2} is more complicated to implement mathematically.
No exact solution exists even for the simplest case of free diffusion with
an absorbing boundary in the underdamped limit, and a set of approximation
schemes has been devised~\cite{burs81}. However, for sufficiently steep
barriers we expect that the solution with free boundary condition is a
good approximation to the absorbing boundary case~\cite{zhan07}. The
approximation will be checked by comparing analytical results with
numerical simulations.

In this work we will consider the initial velocity to be thermalized. This
can be accounted for by integrating $\mathcal{P}$ over all possible
initial velocities. In addition, as a transition path with free boundaries
is defined by the fact that the particle crosses the $x_0$ boundary with
any velocity, its final velocity has to be integrated over. We thus define
\begin{equation}
\label{eq:P}
\text{P}(x, t |-x_0, 0)
\equiv \int\limits_{-\infty}^\infty \dd v
\int\limits_{-\infty}^\infty \dd v_0 \; 
p_\text{eq}(v_0)
\mathcal{P}(x,v, t|-x_0, v_0, 0),
\end{equation}
where 
\begin{equation}
\label{eq:peq}
p_\text{eq}(v) = \sqrt{\frac{m}{2 \pi k_B T}} 
\,\,\,
e^{ -\displaystyle\frac{ m v^2}{2k_BT} } 
\end{equation}
is the Maxwell velocity distribution for a system in equilibrium at
temperature $T$, as we have assumed that the particles
spend sufficiently long times in a given well, so that their positions
and velocities follow an equilibrium distribution. 

To connect the probability of finding the particle at position $x$
at a time $t$ to the TPT distribution, we introduce the 
\textit{absorption function} 
$\QA$ obtained by integrating the function $\text{P}(x, t | -x_0, 0)$
obtained from Eq.~\eqref{eq:P} in the domain $x > x_0$:
\begin{equation}
\label{eq:Qa}
\QA(t) \equiv \int\limits_{x_0}^{\infty} 
\dd x \; \text{P}(x, t | -x_0, 0),
\end{equation} 
which counts all trajectories, originating in $-x_0$, that have already
crossed the boundary $x_0$ at time $t$. Obviously, all these trajectories
have a transition path time smaller than $t$, and thus $\QA(t)$ is
proportional to the probability that the TPT is smaller than $t$.

The difference $\QA(t + \Delta t)-\QA(t)$ is thus equal to the 
fraction of
trajectories that cross the boundary $x_0$ in the interval $[t,t+\Delta
t]$, hence the TPT distribution can be approximated as
\begin{equation}
\label{eq:Ptpt}
p_\text{TP}(t) \approx C \frac{d \QA(t)}{dt},
\end{equation} 
where $C$ is a normalization constant. This expression is approximate
as we are not imposing the appropriate absorbing boundary conditions.
When using free boundary conditions the left-hand side of Eq.~\eqref{eq:Ptpt}
counts also paths with multiple crossings at $-x_0$ and $x_0$, which,
strictly speaking, are not transition paths. However, in the high 
barrier limit, these paths become exceedingly rare and $\QA$ is a
good approximation to the absorbing boundaries case. Finally,
the normalization constant can be obtained from
\begin{eqnarray}
\label{eq:norm}
\int_0^{+\infty} p_\text{TP}(t) dt &=& C \left[ \QA(+\infty) - \QA(0) \right]
\nonumber\\
&=& C \QA(+\infty) = 1,
\end{eqnarray} 
where we have used $\QA(0)=0$, from Eq.~\eqref{eq:Qa}.

\section{TPT distribution for parabolic barrier}
\label{sec:barrier}

We consider here a parabolic barrier centered in $x=0$, which corresponds to
a repulsive linear force
\begin{equation}
\label{U}
F(x) =  K x.
\end{equation}
For this system the Langevin equation~\eqref{eq:Lang} is linear and can
thus be solved. Denoting the solution by $x(t)$, the full propagator
can be written as
\begin{equation}
\mathcal{P}(x, v, t| x_0,v_0,0)= \langle \delta(x-x(t)) \delta (v-\dot x(t)) 
\rangle.
\end{equation}

As we saw in the previous section, the quantity of interest in
Eq.(\ref{eq:P}) is
\begin{equation}
\text{P}(x, t| x_0,0)= \int\limits_{-\infty}^\infty \dd v
\,\dd v_0 \; p_\text{eq}(v_0) 
\langle \delta(x-x(t)) \delta (v-\dot x(t)) \rangle
\end{equation}

The integration over the velocity $v$ is trivial (see details
in Appendix \ref{sec:app1}) and we find 
\begin{align}
\label{eq:distribution}
	& \text{P}(x, t| x_0,0) \notag \\[2 pt]
	& \;\; = \int\limits_{-\infty}^\infty \dd v_0
\; p_\text{eq}(v_0) 
\frac{1}{\sqrt{ 2 \pi \phi^2(t)}}
e^{-\displaystyle\frac{(x-X_{v_0}(t))^2}{2 \phi^2(t)}}.
\end{align}
Here $X_{v_0}(t)$ is the solution of the deterministic equation of motion
for a particle originating in $x_0$ and with initial velocity $v_0$ (see
\eqref{app:xv0}). The variance $\phi^2(t)$, given in Eq.~\eqref{app:phi2},
does not depend on the initial conditions $x_0$ and $v_0$. It vanishes
at short times, while it diverges exponentially at large times.

The next step is the integration in the initial velocities. Since $v_0$
enters linearly in $X_{v_0}$ and $p_\text{eq}(v_0)$ is Gaussian,
Eq.~\eqref{eq:P} boils down to a Gaussian integral which can be easily
performed (for details see Appendix~\ref{sec:app1}). The result is
\begin{equation}
\label{eq:distribution_maxwell}
\text{P}(x, t| x_0, 0) = 
\frac{1}{\sqrt{2 \pi \sigma^2(t)}} \,
{e^{-\displaystyle\frac{\left(x-X_0(t)\right)^2}{2 \sigma^2(t)}}},
\end{equation}
where $X_0(t)$ is the deterministic solution of the equation of motion
for a particle with vanishing initial velocity $v_0=0$. The resulting
distribution \eqref{eq:distribution_maxwell} is again Gaussian, but with
$\sigma^2(t) = \phi^2(t) + \psi^2_v (t)$, i.e. the variance is larger
than that of the distribution \eqref{eq:distribution}. Here $\psi^2_v
(t)$ indicates the contribution obtained from integrating over the
initial velocities (see Eq.~\eqref{app:def_psiv}).

The final step is the calculation of $\QA$, for which we get
(details in Appendix~\ref{sec:app1})
\begin{equation}
	\label{eq:QA_quadr}
	\QA(t) = \frac{1}{2}\left(1 - \erf(G(t)) \right),
\end{equation}
where the error function is defined as 
\begin{equation}
\erf(x) \equiv \frac{2}{\sqrt{\pi}} \int\limits_{0}^{x} \dd u \; e^{-u^2}
\label{eq:erf}
\end{equation}
and
\begin{equation}
G(t) \equiv \frac{x_0 - X_0(t)}{\sqrt{2 \sigma^2(t)}}.
\label{eq:G}
\end{equation}
The full expressions for $X_0(t)$ and $\sigma(t)$ are given in the
Appendix~\ref{sec:app1}, Eqs.~\eqref{app:X0} and~\eqref{app:sigma2}.
Using Eqs.~\eqref{eq:Ptpt}, \eqref{eq:norm} and the calculation of the
normalization constant given in~\eqref{app:normalization}, we arrive at
the following expression for the TPT distribution:
\begin{equation}
\label{eq:PTPT_quadr}
p_\text{TP}(t) = 
-\frac{2}{\sqrt{\pi}}\frac{G'(t)e^{-G^2(t)}}{1-\erf(\sqrt{\beta E})}.
\end{equation}
where $G'\equiv dG/dt$ is the time derivative of $G(t)$, $\beta=1/k_BT$
and $E=Kx_0^2/2$ is the barrier height.

\begin{figure}
\includegraphics[width = 0.45 \textwidth]{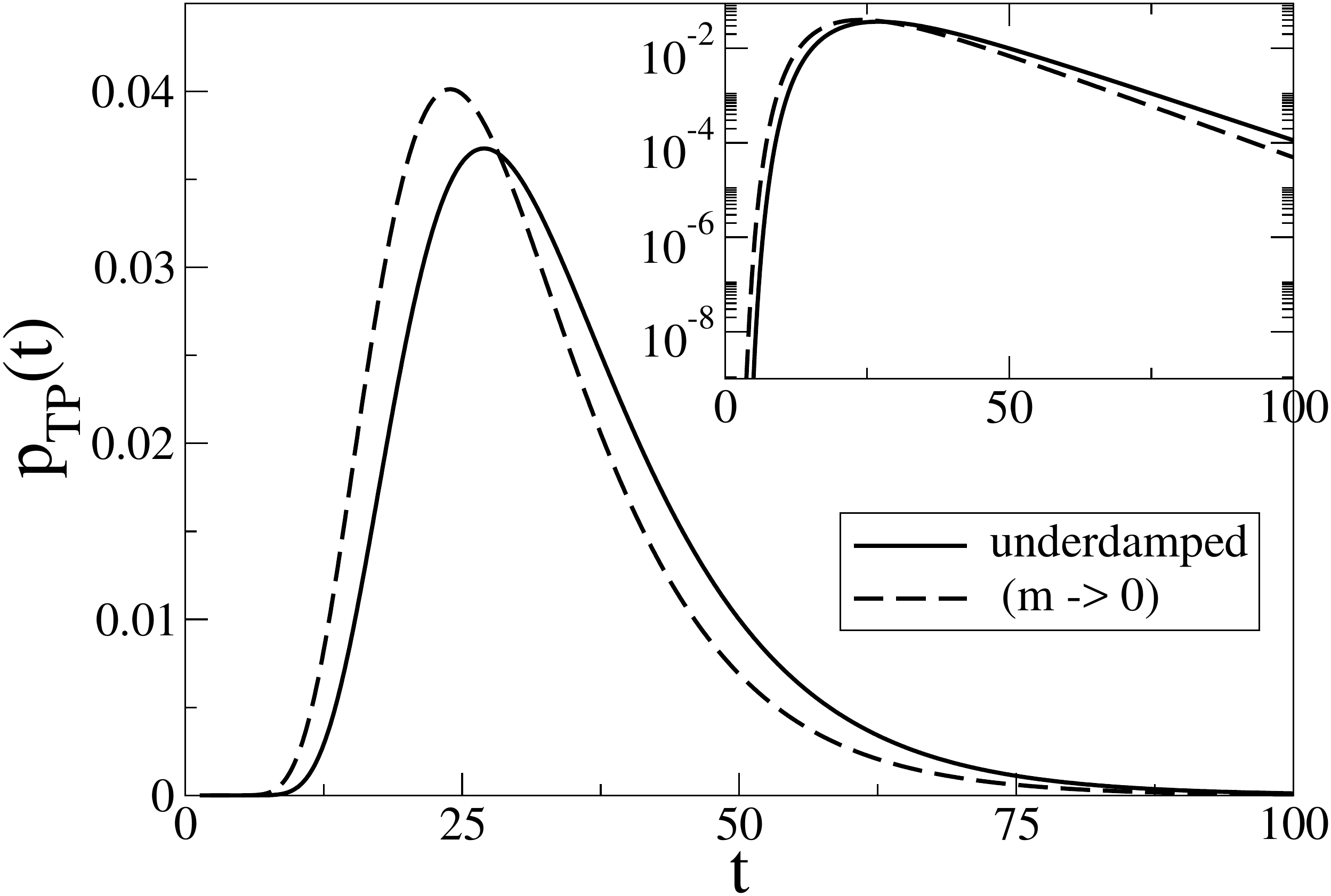}
\caption{Transition path times distribution as obtained from
Eq.~\eqref{eq:PTPT_quadr} for an underdamped particle (solid line) and
for the overdamped limit (dashed line) obtained by setting $m \to 0$.
The characteristic behavior at short and long times is summarized in
Table~\ref{table1}. The parameters used are, in dimensionless units,
$x_0 = -10$, $K = 0.1$, $\gamma = 1$, $m = 1$ and $\kB T = 1$ corresponding
to $\beta E = 5$. Inset: the same data in semi-logarithmic scale to
emphasize the exponential decay of the distribution at long times.}
\label{fig:TPT}
\end{figure}

The solid line in Fig.~\ref{fig:TPT} shows a plot of the TPT distribution
for an underdamped particle as given by Eq.~\eqref{eq:PTPT_quadr}.
This distribution vanishes at short times, with a leading singular
behavior due to the divergence of $G(t)$ in the limit $t \to 0$. From
Eqs.~\eqref{app:X0} and \eqref{app:sigma2} one finds at short times
\begin{equation}
p_\text{TP}(t) {\underset{t \to 0}{\sim}} e^{-G^2(t)} = 
\exp\left( -\frac{2 \beta mx_0^2}{t^2} \right).
\label{eq:TPT_ss}
\end{equation}

This result can be understood from the ballistic motion of particles
starting from $-x_0$.  At short times the effect of the parabolic
potential can be neglected and particles follow a free motion $x(t)
\approx - x_0 + v t$.  The variance is then given by:
\begin{equation}
\sigma^2 (t) = \langle \left( x(t) + x_0 \right)^2 \rangle \approx 
\langle v^2 \rangle t^2 = \frac{k_BT}{m} \,t^2,
\end{equation}
where we have used the equipartition theorem $m \langle v^2 \rangle =
k_BT$.  Plugging in this result in Eq.~\eqref{eq:G} and recalling that
$X_0(t) \approx -x_0$ at short times (Eq.~\eqref{app:X0}) we obtain:
\begin{equation}
G^2(t) = \frac{\left( x_0 - X_0(t)\right)^2}{2 \sigma^2(t)} 
\,
{\underset{t \to 0}{\sim}}
\,\,
\frac{2x_0^2}{\sigma^2(t)} = \frac{2mx_0^2}{k_BT} \, t^{-2},
\end{equation}
which explains the result of Eq.~\eqref{eq:TPT_ss}.

At long times $G(t)$ converges to a finite constant, since
both the numerator and denominator in Eq.~\eqref{eq:G} diverge
at the same rate. The calculation gives $G^2(t) \to \beta E$
(Eq.~\eqref{app:G2lim}). However the derivative $G'(t)$ vanishes
exponentially at large $t$ (Eq.~\eqref{app:Gplim}) as
\begin{equation}
G'(t) 
\,
{\underset{t \to \infty}{\sim}}
\,
e^{-\lambda_+ t}
\end{equation}
with 
\begin{equation}
\lambda_+ = \frac{-\gamma+\sqrt{\gamma^2+4Km}}{2m}
\end{equation}
a characteristic rate of the process.  The limiting behaviors of $G^2(t)$
and $G'(t)$ are summarized in Table~\ref{table1}.

\begin{table}[t]

\begin{tabular}{|c|c|c|c|}
\hline
&rates& short times & long times \\

\hline
&&&\\
{\multirow{3}{*}{\rotatebox[origin=c]{90}{underd.}}}
& $\lambda_{\pm} \equiv \displaystyle{\frac{-\gamma\pm \sqrt{\Delta}}{2m}}$ 
& 
$G^2(t) \sim \displaystyle\frac{2 \beta m x_0^2}{t^2}$
& $G'(t) \sim - \sqrt{\beta E} \, \times $\\ 
& $\Delta = \gamma^2 + 4Km$
&
&
$\displaystyle{\frac{2 \sqrt{\Delta}}{\gamma+\sqrt{\Delta}}} \lambda_+ e^{-\lambda_+ t}$
\\[6 pt]
&&&\\

\hline
&&&\\
\rotatebox[origin=c]{90}{overd.}
& $\Omega \equiv \displaystyle{\frac{K}{\gamma}}$ & 
$G^2(t) \sim \displaystyle\frac{\beta \gamma x_0^2}{t} $
& {$G'(t) \sim - \sqrt{\beta E} \, \Omega e^{-\Omega t}$} \\
&&&\\

\hline
\end{tabular}
\caption{Summary of the asymptotic forms of $G(t)$ and $G'(t)$
determining the behavior of the TPT distribution~\eqref{eq:PTPT_quadr}
at short and long times. Both the underdamped and overdamped regimes
are given. (Note: short and long times correspond to the limits $\Omega
t \ll 1$, $|\lambda_\pm t|\ll 1$ and $\Omega t \gg 1$, $|\lambda_\pm
t|\gg 1$, respectively).}
\label{table1}
\end{table}

\subsection*{The overdamped regime}

The overdamped limit is discussed in Appendix~\ref{app:overdamped}
In this case the function $G(t)$ has a particularly simple expression:
\begin{equation}
\label{G2}
G^2(t) = \beta E \, \frac{1+\exp(-\Omega t)}{1-\exp(-\Omega t)},
\end{equation}
where $\Omega \equiv K/\gamma$ defines the characteristic rate
of the overdamped system. This distribution was derived 
in~\cite{zhan07}. The short time behavior is:
\begin{equation}
G^2(t) \approx \frac{2\beta E}{\Omega t} = 
\frac{\beta \gamma x_0^2}{t},
\label{Goverd}
\end{equation}
as expected from diffusive behavior of the particles we have at short
times $2 \sigma^2 \approx 4 D t$, while the numerator $(x_0 - X_0(t))^2
\approx 4 x_0^2$. Using the Einstein relation $D = k_B T/\gamma$ we
recover Eq.~\eqref{Goverd}.  At long times $G^2(t) \to \beta E$ and
\begin{equation}
G'(t) \approx - \sqrt{\beta E}\, \Omega e^{-\Omega t}
\end{equation}

\subsection*{Comparing the overdamped and underdamped cases}

Figure~\ref{fig:TPT} shows a comparison of the underdamped
TPT distribution (solid line) with the overdamped one, obtained by
setting the mass term to zero (dashed line). Note that the values
of the parameters chosen, given in the figure caption, correspond
$\lambda_+=0.092$ while $\Omega=0.1$. Hence the differences between
the two cases is expected to be rather small, as shown indeed in
Fig.~\ref{fig:TPT}. The overdamped limit $m \to 0$ leads to a shift
of the distribution to shorter timescales compared to the underdamped
limit. Therefore inertia globally slows down the transition paths
dynamics in analogy to the slowing down of Kramers' times discussed above
(Eqs.~\eqref{eq:Kramers2} and~\eqref{eq:Kramers}).  Note that at short
time scales there is an opposite behavior with more faster crossings
events in the underdamped regime (Eq.~\eqref{eq:TPT_ss}) compared
to the overdamped regime (Eq.~\eqref{Goverd} implies $p_\text{TP}(t)
\sim \exp(\beta \gamma x_0^2/t)$). This behavior however involves a
neglegible fraction of transition paths as shown in Fig.~\ref{fig:TPT}
(a second crossing between dashed and solind line takes place at short
time scales, this is expected at times $1/|\lambda_-|\approx 0.92$).

\begin{figure}[t]
\includegraphics[width = 0.45 \textwidth]{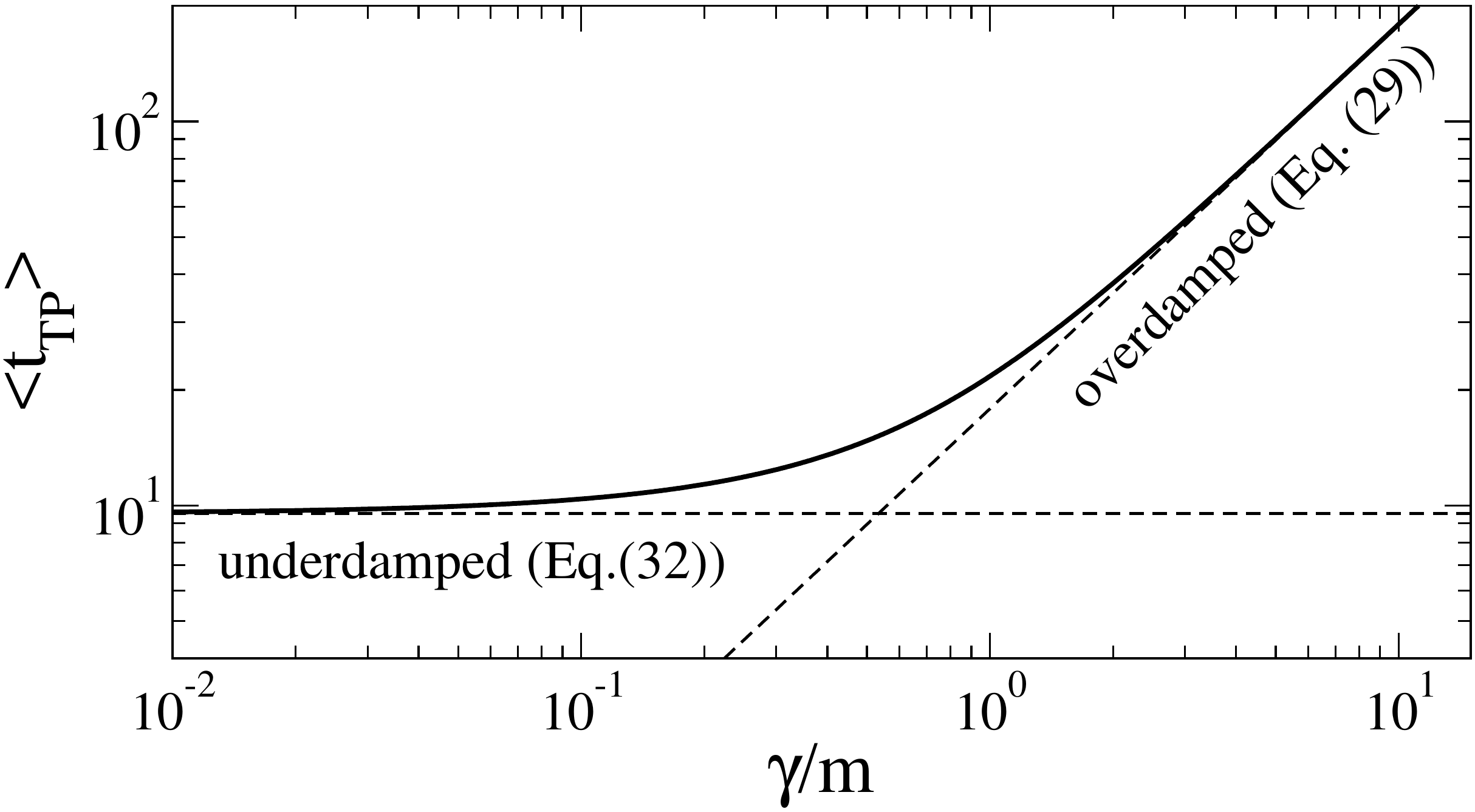}
\caption{Solid line: average TPT plotted as a function of $\gamma/m$
in the larger barrier limit obtained from a numerical estimate of
the integral in Eq.~\eqref{av_TPT}, with parameters $x_0=2$, $K = 
10$, $m = 50$, $\kB T = 2$. Dashed lines: limiting values for
$\langle t_\text{TP} \rangle $ in the over- and underdamped regime give
by Eqs.~\eqref{TPT_over} and \eqref{TPT_under}, respectively.}
\label{fig:av_t}
\end{figure}

\section{The average TPT}
\label{sec:av_TPT}

The average value of TPT is given by
\begin{equation}
\label{av_TPT}
\langle t_\text{TP}  \rangle= \int_0^{+\infty} dt \, t \, p_\text{TP} (t) =
\frac{\displaystyle\int_{\sqrt{\beta E}}^{+\infty} t(G) \, e^{-G^2} dG}
{\displaystyle\int_{\sqrt{\beta E}}^{+\infty} e^{-G^2} dG}
\end{equation}
where we have rewritten the integral using a change of variable $G'
dt = dG$.  From the analysis of the previous section we have seen that
$G(t)$ is a monotonic decreasing function of $t$ and $\sqrt{\beta E}
\leq  G \leq +\infty$. 

We now present the results for the behaviour of the average TPT for
large barrier. This asymptotic behavior is dominated by the behavior
of $t(G)$ close to the lower bound $G \gtrsim \sqrt{\beta E}$, where
both underdamped and overdamped TPT distributions decay exponentially
(see Table~\ref{table1}). The details of the calculations can be found
in the Appendix.

In the overdamped regime, Eq. \eqref{G2} can be inverted easily to yield
$t$ as a function of $G$, and we obtain asymptotically for large barrier
$\beta E \gg 1$
\begin{equation}
\label{TPT_over}
\langle t_\text{TP}  \rangle \approx \frac{\gamma}{K} \log( 2 e^{C} \beta E),
\end{equation}
where $C \approx 0.577 215$ is the Euler-Mascheroni constant. This result
coincides with that previously obtained by A. Szabo~\cite{chun09}. 

In the underdamped regime, Eq. \eqref{eq:G} cannot be inverted to yield 
$t$ as a function of $G$. However, using the asymptotic form of $G$, we show in 
the appendix that the large barrier limit of the average TPT is given by
\begin{equation}
\label{TPT_under}
\langle t_\text{TP}  \rangle \approx \frac{1}{\lambda_+}\left(  \log \beta E +A 
\right),
\end{equation}
where $A$ is a constant independent of the barrier height given by
\be
A= \log \left( \frac {4 \sqrt \Delta}{\gamma + \sqrt \Delta} e^C \right),
\label{defA}
\ee
where as above, $C$ is the Euler-Mascheroni constant. The other parameters
are defined in the appendix and in Table~\ref{table1}.  Note that
this formula is quite general, as it yields back the overdamped case
(\ref{TPT_over}) in the limit of vanishing mass. In the limiting case of
$\gamma \to 0$ Eqs.~\eqref{TPT_under} and~\eqref{defA} become
\be
\langle t_\text{TP} \rangle = \sqrt{\frac{m}{K}} \, \log \left( 4 e^C \beta E\right)
\ee

Figure~\ref{fig:av_t} shows a plot of $\langle t_\text{TP} \rangle$
obtained by numerical integration of the probability distribution of TPT
as a function of the ratio $\gamma/m$. At very high friction $\gamma^2 \gg
4Km$ the calculation reproduces the overdamped limit~\eqref{TPT_over}. At
low friction $\gamma^2 \ll \Delta \approx 4Km$ (\ref{TPT_over}) converges
to a friction independent limit, shown by the dashed horizontal line in
Fig.~\ref{fig:av_t}.  Some words of caution are necessary when dealing
with the low friction limit. It is well known~\cite{hang90} that the
theory developed here is not valid when the friction becomes arbitrary
small, but one still needs $\gamma \beta E \gg \sqrt{K/m}$. For friction
below this limit the particles do not thermalize at the bottom of one well
and the use of Eqs.~\eqref{eq:P} and~\eqref{eq:peq} is not justified. As
for the Kramers' time one expects an increase in the average TPT at very
low friction (known as Kramers' turnover~\cite{hang90}).

\section{Numerical Results}
\label{sec:numerics}

To assert the range of validity of the analytical calculations, numerical
simulations were performed. This was done by numerically integrating
the Langevin equation \eqref{eq:Lang} using the algorithm developed by
Vanden-Eijnden and Ciccotti\cite{vde06}.

\begin{figure}[t]
\includegraphics[width = 0.45 \textwidth]{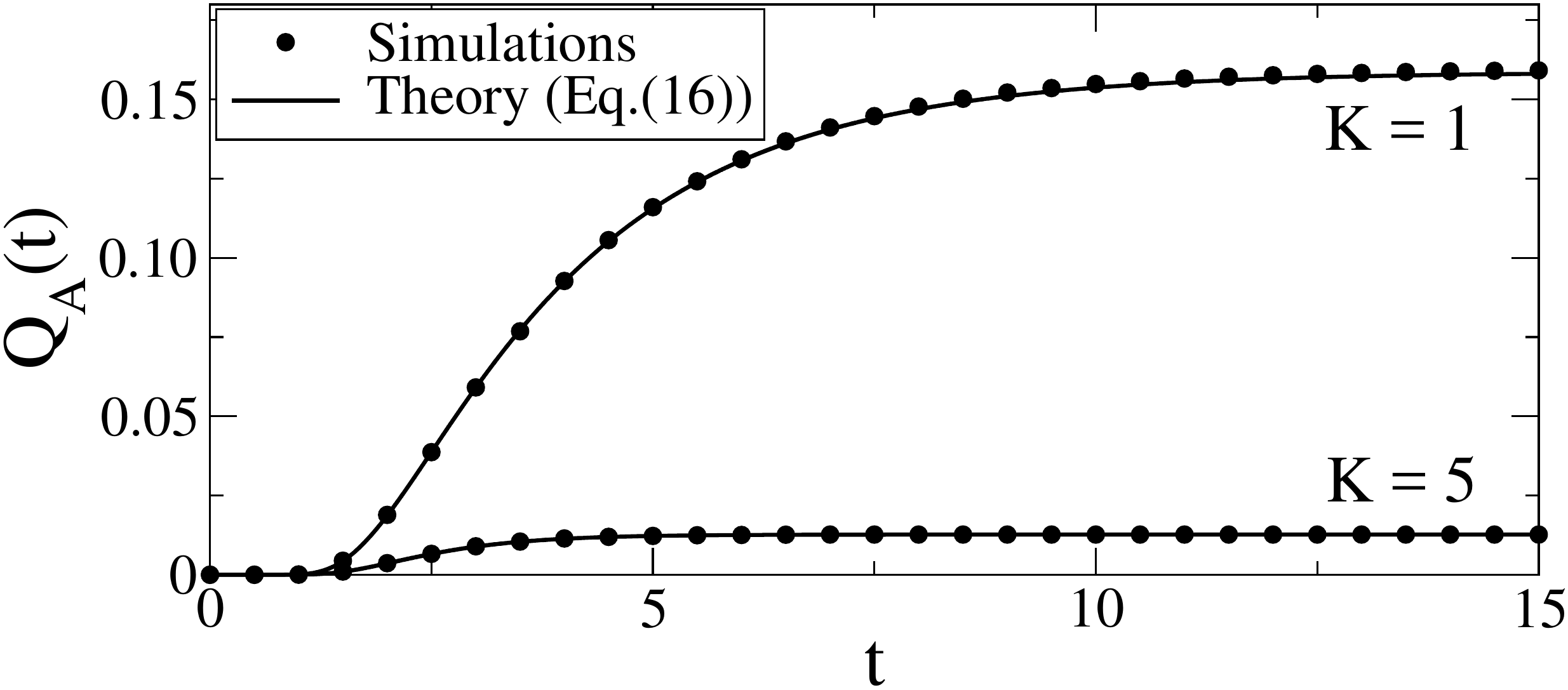}
\caption{Comparison between analytical and numerical estimates of
absorption probabilities $\QA$ for a thermalized set of particles
started at $x = -x_0$. The parameters are $x_0 = 1$, $m = 1$, $\gamma =
1$ and $T = 1$. Two values of $K$ have been simulated.}
\label{fig:Q_sim}
\end{figure}

\subsection*{The absorption probability $\QA$} 

We run a first set of simulations using free boundary conditions
and calculated the absorption probability $\QA(t)$. As the analytical
expression~\eqref{eq:Qa} was also obtained with free boundary conditions,
analytical and numerical results must coincide. The runs are used to test
the accuracy of the integration scheme. To obtain $\QA(t)$ from simulations
an ensemble of thermalized particles was put at the initial position $x =
-x_0$ and subsequently evolved through time. At regular time intervals
the number of particles in the region $x(t) > x_0$ was counted, which is,
after normalizing by dividing through the total number of particles,
a direct measure of $\QA$.  The result of the simulation is shown on
Fig. \ref{fig:Q_sim} for two different values of the barrier curvature
$K=1$ and $K=5$. At long times the absorption probability saturates to
a constant, which decreases as $K$ is increased since a smaller fraction
of particles cross the barrier as this becomes steeper. 

\begin{figure}[t]
\includegraphics[width = 0.45 \textwidth]{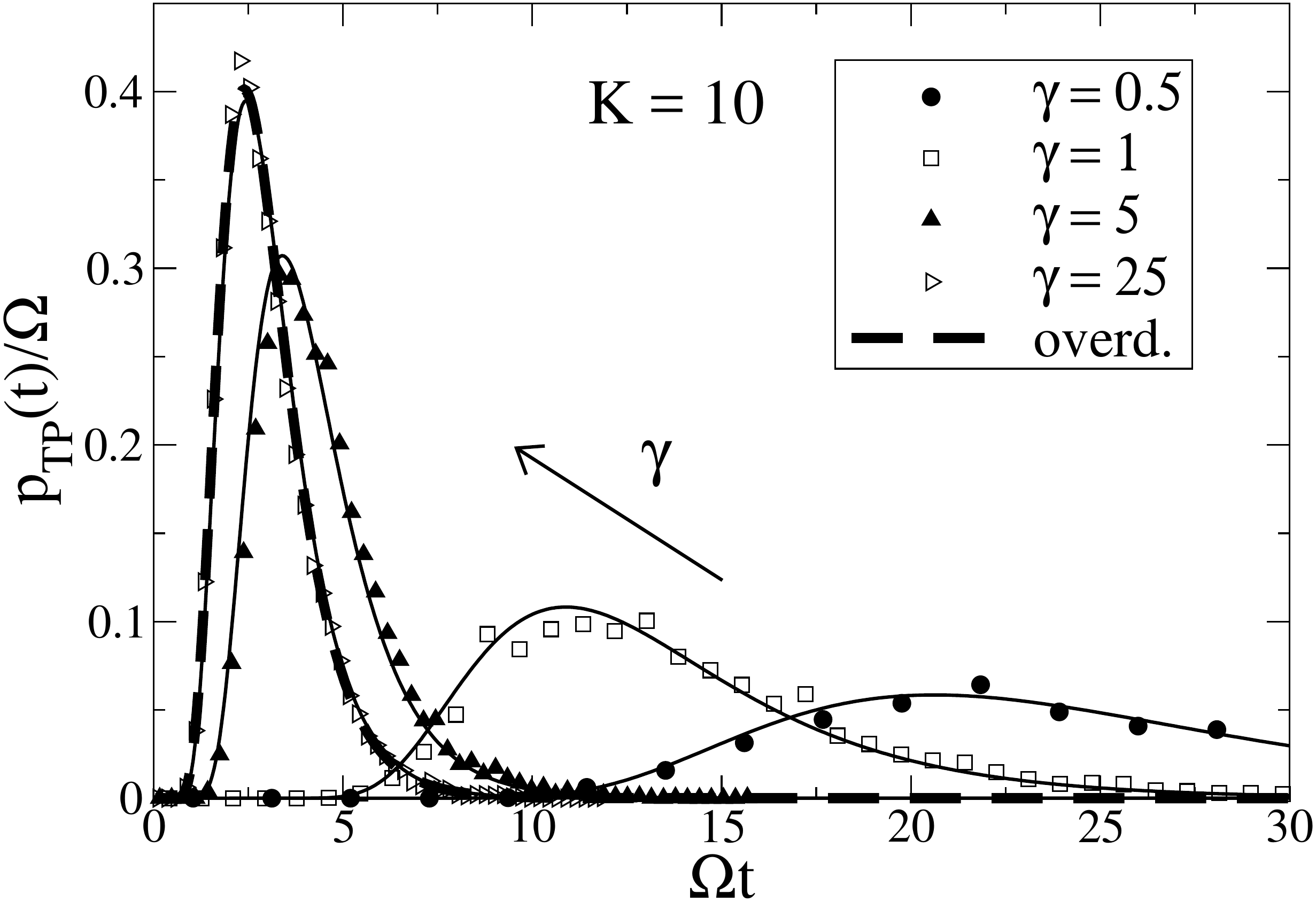}

\includegraphics[width = 0.45 \textwidth]{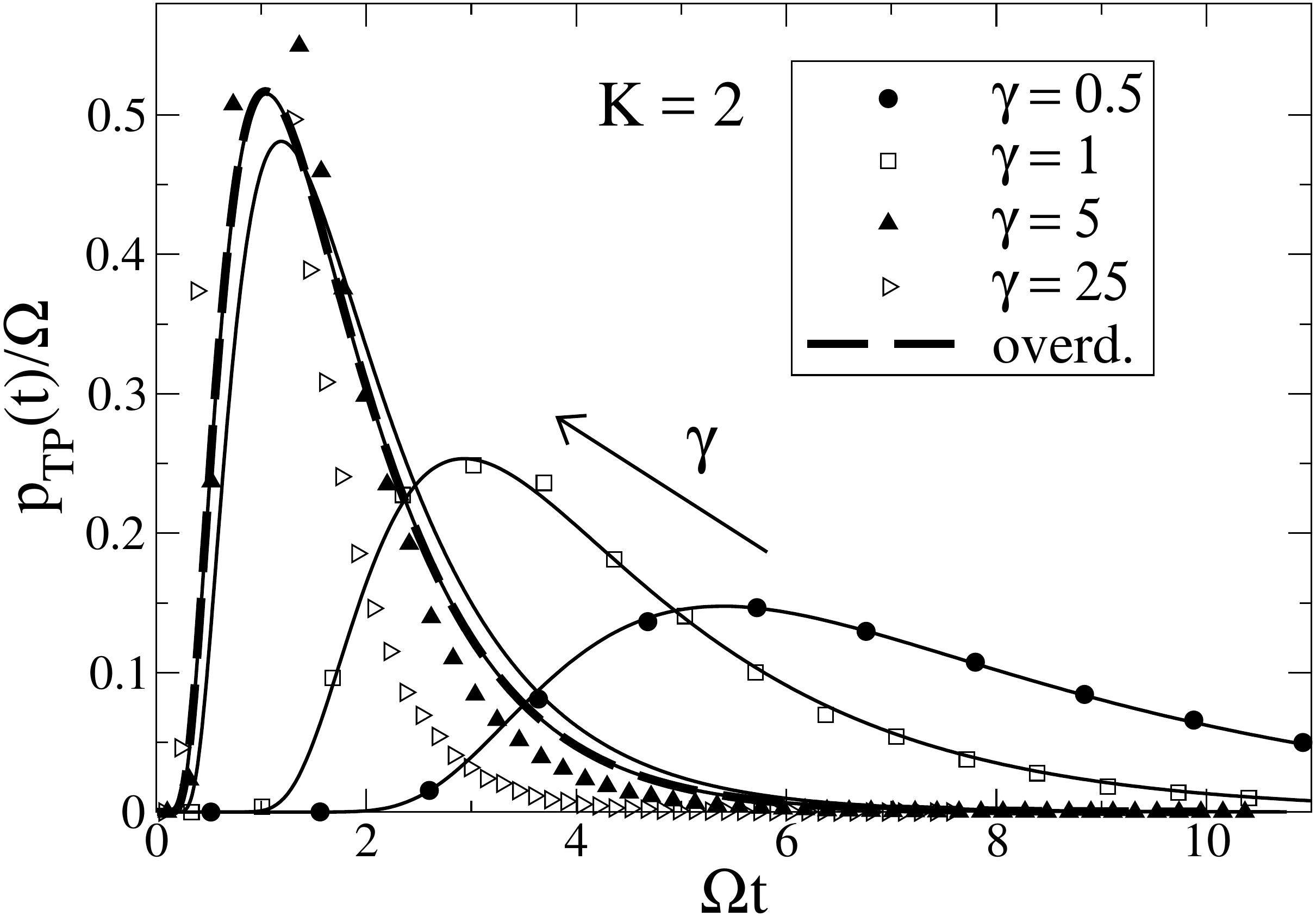}
\caption{Solid lines: underdamped TPT distributions obtained from
Eq.~\eqref{eq:PTPT_quadr} for various $K$ and $\gamma$. Dashed lines:
overdamped TPT distributions. Note that the solid line and dashed line
coincide for the highest friction data. Symbols: TPT distributions from
simulations. The other parameters are $x_0=1$ and $k_BT=1$. Deviations are
observed between theory and experiments for $K=2$ and high friction (see
discussion in the text).}
\label{fig:ptpt}
\end{figure}

\subsection*{The TPT distribution}

We computed next the TPT distribution from simulations imposing absorbing
boundary conditions in $x_0$.  In order to obtain sufficient statistics
a forward flux sampling scheme~\cite{allen2005} was employed. In this
scheme the transition path is created step by step rather than at
once. The dynamics of the particles still starts at $-x_0$, but instead
of waiting until one of them hits $x_0$, we evolve them until they
hit $-x_0 + \epsilon$ (where $\epsilon $ is a small, positive number),
where we store the physical state of the particle. On the other hand,
whenever a particle ventures into $x \leq -x_0$, we discard it and
restart the simulation for that particle. This procedure is  repeated
until a representative ensemble at $-x_0 + \epsilon$ is generated. In
the next step, we evolve the particles from $-x_0 + \epsilon$ to $-x_0 +
2 \epsilon$, where we sample the initial conditions for this step from
the ensemble generated by the particles that reached $-x_0 + \epsilon$
in the previous step. Here again, the physical state is stored whenever
$-x_0 + 2 \epsilon$ is reached, or we discard and restart when $x \leq
-x_0$. These steps are repeated until $x_0$ is reached.

Figure~\ref{fig:ptpt} shows plots of TPT distribution obtained for various
values of the parameters and compares simulations (symbols) with the
theory from Eq.~\eqref{eq:PTPT_quadr} (solid lines). The dashed line is
the overdamped distribution which overlaps with Eq.~\eqref{eq:PTPT_quadr}
for the highest value of the friction ($\gamma=25$). The data are
plotted in rescaled time unit $\Omega t$, where $\Omega=K/\gamma$ is
the characteristic rate of the overdamped case. The two plots correspond
to two values of the barrier stiffness $K=2$ and $K=10$, while $x_0=1$
and $k_BT=1$. For each set we plot four different values of $\gamma$
(where the arrow indicates the direction of increasing $\gamma$).

For the steeper barrier ($K=10$) there is an excellent agreement between
theory and simulations for all values of $\gamma$. In this case there is
practically no difference between the free boundary conditions (theory)
and the absorbing boundary conditions (simulations). This is in line
with previous studies of the overdamped distributions~\cite{zhan07}.
Deviations are instead observed in the $K=2$ data, which are clearly
visible at high friction, close to the overdamped limit. Here the
analytical calculation overestimates the TPT, leading to a broader
distribution compared to the simulations. This is because the theory with
free boundary conditions wrongly counts as TP also those trajectories
with multiple crossings at the boundaries and these trajectories lead to
high TPT. At low friction and $K=2$, however, theory and simulation match
again. This agreement can be understood as follows. As highlighted in the
calculations of Section~\ref{sec:barrier} there are two contributions
to the noise. Firstly, there is an intrinsic noise, contributing to
the variance $\phi^2(t)$ in Eq.~\eqref{eq:distribution}. Secondly, the
thermalization over different initial velocities leads to an additional
contribution $\psi^2_v(t)$ to the variance, so that the total variance is
$\sigma^2(t) = \phi^2(t) + \psi^2_v(t)$ (see Eq.~\eqref{app:def_psiv}),
A simple analysis shows that $\phi^2(t)$ vanishes at low $\gamma$, whereas
$\psi^2_v(t)$ converges to a non-vanishing constant.  At small friction
the trajectories are weakly influenced by thermal noise, particles will
follow closely the deterministic trajectory, hence in the free boundary
case multiple crossings of the boundary at $x_0$ will be rare. This is
the reason of the agreement between theory and simulations. Note that the
predominant contribution to the width of the TPT distribution comes in
this limit from $\psi^2_v(t)$, i.e. the distribution of initial velocities.

\section{Discussion}
\label{sec:discussion}

Conformational transitions molecular systems between two different states
are governed by two time scales. The Kramers time corresponds to the
typical time spent in a given conformation, while the transition path time
characterizes the actual duration of the transition.  Transition path
times, which have been measured in protein and nucleic acids folding
experiments during the past decade~\cite{chun09,neup12,true15,neup17},
can be a few orders of magnitudes shorter than Kramers' times.

In this paper we have analyzed the TPT distribution of a simple one
dimensional stochastic particle undergoing Langevin dynamics and crossing
a parabolic barrier. We focused in particular to the underdamped case,
thus extending previous analysis~\cite{zhan07} in which inertial terms
were neglected. As the barrier is parabolic, the associated Langevin
equation is linear and hence exactly solvable. The inclusion of inertia
makes the calculations more complex than in the overdamped limit, but
an analytical form of the TPT distribution can still be obtained. This
solution is not exact as it does not use the appropriate absorbing
boundary conditions, but it approximates very well the numerical
simulations for steep barriers.

In general inertia slows down the barrier crossing process. The
main properties of the TPT distribution have been summarized in
Table~\ref{table1}.  The distribution has an essential singularity at
short times, which is however of different nature in the overdamped and
underdamped cases.  At long times the distribution vanishes exponentially
in both cases. Differently from the Kramers' time, which is characterized
by an exponential dependence on the barrier height $E$, the average
TPT in the overdamped limit scales logarithmically $\langle t_\text{TP}
\rangle \sim \log (\beta E)$, where $\beta$ is the inverse temperature.
We have shown here that the logarithmic dependence also holds in the
underdamped limit and have calculated the prefactor.

We conclude by remarking that, although in the analysis of the dynamics
of molecular systems the overdamped limit is usually considered, owing to
the short duration of TPT it is possible that inertial effects influence
the barrier crossing dynamics. The availability of an analytical closed
form of the TPT distribution in the underdamped case may indeed be useful
for the analysis of future experiments.

\bigskip

\begin{acknowledgments}
One of us (HO) would like to thank Michel Bauer for useful
discussions.
\end{acknowledgments}

%
%
%
%

\appendix

\begin{widetext}

\section{Details of the derivation of the TPT distribution}
\label{sec:app1}
\setcounter{equation}{0}
\renewcommand{\theequation}{A\arabic{equation}}

We give here all the details of the calculations, which were just briefly
outlined in Section \ref{sec:TPT_gen}.

\subsection{The probability distribution}

The solution of the second order linear differential equation
\begin{equation}
m\ddot{x} = Kx -\gamma \dot{x}  + \eta(t).
\end{equation}
with initial conditions $x(0)=-x_0$ and $\dot{x}(0) = v_0$ is given by
\begin{equation}
x(t) = \frac{m}{\sqrt{\Delta}} 
\left[  
e^{\lambda_+ t} (\lambda_- x_0 + v_0)  \, - \, 
e^{\lambda_- t} (\lambda_+ x_0 + v_0)  
\right]
+ \frac{1}{\sqrt{\Delta}} \int_0^t \dd \tau \; \left(e^{\lambda_+(t - \tau)} \, - \, 
e^{\lambda_- (t - \tau)} \right) \eta(\tau),
\label{eq:sol}
\end{equation}
where the rate constants are given by:
\begin{equation}
\lambda_{\pm} = \frac{-\gamma \pm \sqrt{\Delta}}{2m} 
\end{equation}
and
\begin{equation}
\Delta \equiv \gamma^2 + 4 m K.
\end{equation}

Note that $\lambda_+ >0$, $\lambda_-<0$ and $\lambda_+ - \lambda_- =
\sqrt{\Delta}/m$. In addition in the high friction limit $\gamma^2 \gg
4mK$ one has
\begin{equation}
\lambda_+ \approx \frac{K}{\gamma} 
\qquad \text{and} \qquad
\lambda_- \approx -\frac{\gamma}{m}
\label{app:lambda_lim}
\end{equation}
where in this limit $\lambda_+$ describes the deterministic dynamics of an
overdamped particle ``sliding down" in a parabolic potential: the equation
of motion is $\gamma\dot{x}=Kx$ and the solution, with initial condition
$x(0)=-x_0$ is $x(t) = -x_0 e^{Kt/\gamma}$. The factor $1/|\lambda_-|$
is the timescale beyond which inertia effects are negligible.

Averaging \eqref{eq:sol} over different noise realizations we obtain 
the solution of the deterministic equation of motion
\begin{equation}
X_{v_0}(t) \equiv \langle x(t) \rangle = \frac{m}{\sqrt{\Delta}} 
\left[
e^{\lambda_+ t} (\lambda_- x_0 + v_0)  \, - \, 
e^{\lambda_- t} (\lambda_+ x_0 + v_0)  
\right]
\label{app:xv0}
\end{equation}
where the subscript $v_0$ indicates the initial velocity (the
solution obviously also depends on the initial position $-x_0$,
however we omit to indicate this explicitly).
Using Eqs.~\eqref{eq:gaussian_white_noise} we obtain for the variance:
\begin{eqnarray}
\phi^2(t) &\equiv& \left\langle (x(t) - X_{v_0}(t))^2 \right\rangle =  
\frac{1}{\Delta} \int_0^t d\tau d\tau' 
\left(e^{\lambda_+(t - \tau)} \, - \, e^{\lambda_- (t - \tau)} \right)
\left(e^{\lambda_+(t - \tau')} \, - \, e^{\lambda_- (t - \tau')} \right)
\langle\eta (\tau) \eta(\tau') \rangle =
\nonumber\\
&=&
\frac{2 \kT \gamma}{\Delta} 
\int_0^t d\tau 
\left(e^{\lambda_+(t - \tau)} \, - \, e^{\lambda_- (t - \tau)} \right)^2
=
\frac{2 \kT \gamma}{\Delta} 
\left[ \frac{e^{2 \lambda_+ t} - 1}{2 \lambda_+} \, -\,
2 \frac{e^{ (\lambda_+ + \lambda_-) t} - 1}{\lambda_+ + \lambda_-} \,+\,  \frac{e^{2
\lambda_- t} - 1}{2 \lambda_-} \right].
\label{app:phi2}
\end{eqnarray}

As $x(t)$ is a sum of gaussian stochastic variables (Eq.~\eqref{eq:sol}),
is itself a gaussian stochastic variable and it is fully characterized
by its average and variance. Therefore, we have
\begin{equation}
\mathcal{P}(x,v,t \,|\, -x_0,v_0, 0) = \frac{1}{\sqrt{2 \pi \phi^2}} \exp 
\left[ -\frac{(x - X_{v_0}(t))^2}{2 \phi^2} \right].
\end{equation}

To perform the integral over initial velocities \eqref{eq:P} it is
convenient to rewrite $X_{v_0}$ into two parts, separating the terms
containing $v_0$
\begin{equation}
X_{v_0} (t) = X_0(t) + \frac{m v_0}{\sqrt{\Delta}} \left(e^{\lambda_+ t} - 
e^{\lambda_- t} \right),
\end{equation}
where:
\begin{equation}
X_0(t) \equiv  \frac{mx_0}{\sqrt{\Delta}} 
\left(  \lambda_- e^{\lambda_+ t}
 \, - \, \lambda_+ e^{\lambda_- t}
\right) ,
\label{app:X0}
\end{equation}
describes the deterministic motion of the underdamped particle with
initial conditions $x(0)=x_0$ and $\dot{x}(0)=0$.

The integral over $v_0$ in \eqref{eq:P} is a shifted gaussian distribution
and the result is:
\begin{equation}
\text{P}(x,t \,|\, -x_0, 0) = \frac{1}{\sqrt{2\pi \sigma^2}} \exp 
\left[ -\frac{ \left(x - X_0(t)\right)^2}{2\sigma^2} \right],
\end{equation}
where
\begin{equation}
\sigma^2 (t) =   \phi^2(t) + \frac{ m  \kT}{\Delta} 
\left(e^{\lambda_+ t} - e^{\lambda_- t} \right)^2 =
\frac{2mk_BT}{\Delta} 
\left[
-\frac{\lambda_-}{2\lambda_+}
\left( e^{2 \lambda_+ t} - 1\right)
-\frac{\lambda_+}{2\lambda_-}
\left( e^{2 \lambda_- t} - 1\right)
+ e^{(\lambda_+ + \lambda_-)t} - 1 
\right]
\label{app:sigma2}
\end{equation}
which shows that the averaging over initial velocities leads to an 
increase in the variance by a factor:
\begin{equation}
\psi_v^2 (t) \equiv \frac{ m  \kT}{\Delta} 
\left(e^{\lambda_+ t} - e^{\lambda_- t} \right)^2
\label{app:def_psiv}
\end{equation}

\subsection{The transition path time distribution}

The next step is the calculation of the absorption function 
\begin{equation}
	\QA(t) \equiv \int_{x_0}^\infty \dd x \; \text{P}(x,t \,|\,- x_0, 0) 
	= \int_{x_0}^\infty \dd x \; \frac{1}{\sqrt{2\pi \sigma^2}} \exp 
	\left[ -\frac{\left(x - X_0 (t)\right)^2}{2 \sigma^2} \right].
\end{equation}
Using the following definition of the error function
\begin{equation}
	\erf (t) \equiv \frac{2}{\sqrt{\pi}} \int_0^t \dd u \; e^{-u^2},
\end{equation}
we can explicitly write the absorption rate as:
\begin{equation}
\QA(t) = \frac{1}{2}\Big(1 - \erf \,(G(t)) \Big),
\end{equation}
where
\begin{equation}
G(t) \equiv \frac{x_0 - X_0(t)}{\sqrt{2 \sigma^2(t)}}.
\end{equation}
To obtain the appropriate normalization we need to calculate the $t
\to \infty$ behavior of $G(t)$.  One finds $G^2(t) \to \beta E$ (see
Eq.~\eqref{app:G2lim}), where $\beta=1/k_BT$ is the inverse temperature
and $E\equiv Kx_0^2/2$ the potential barrier that the particle needs
to overcome.  From the previous results we obtain:
\begin{equation}
\lim_{t \to \infty} \QA(t) = 
\frac{1}{2} \lim_{t \to \infty} \left( 1- \erf \, (G(t)) \right) =
\frac{1}{2} \left( 1- \erf \, (\sqrt{\beta E}) \right)
\label{app:normalization}
\end{equation}
The TPT distribution is then given by the derivative of $\QA$ with respect to 
time:
\begin{equation}
p_{TP}(t) 
= \frac{1}{\QA(\infty)}\frac{\dd \QA(t)}{\dd t}
=\frac{1}{2 \QA(\infty)} \frac{\dd}{\dd t} \left( -\erf\left( G(t) 
\right) + 1 \right) 
= -\frac{2}{\sqrt{\pi}}\frac{G'(t)e^{-G^2(t)}}{1-\erf(\sqrt{
\beta E})}.
\end{equation}
where $G'(t) \equiv dG/dt$.

\subsection{Asymptotic behavior}

At short times $|\lambda_\pm|t \ll 1$, we can expand the exponentials
in the above expression for small values of the arguments. Expanding the
exponential to lowest orders we find
\begin{equation}
\sigma^2 (t) \approx \frac{2mk_BT}{\Delta}  \,
\left[ -\lambda_-\lambda_+ - \lambda_+\lambda_- + \frac{\left(
\lambda_+ + \lambda_-\right)^2}{2}
\right] t^2 = \frac{2mk_BT}{\Delta} \frac{\left(
\lambda_+ - \lambda_-\right)^2}{2} \, t^2
= \frac{k_B T}{m} t^2
\label{app:sigma_shortt}
\end{equation}
which is the expected result from the equipartition theorem, as discussed
in Section~\ref{sec:barrier}.  At long times (recall that $\lambda_-<0$
and $\lambda_++\lambda_-<0$) one has
\begin{equation}
\sigma^2 (t) \approx - \frac{mk_B T \lambda_- }{\Delta \lambda_+}  \,
e^{2 \lambda_+ t} \left( 1+ {\cal O}\left(e^{-2 \lambda_+ t}\right)\right)
\label{app:sigma_longt}
\end{equation}
and
\begin{equation}
\left( x_0 - X_0(t)\right)^2 \approx
\left( x_0 - \frac{mx_0}{\sqrt{\Delta}} \lambda_- e^{\lambda_+ t} \right)^2 \approx
\frac{m^2x_0^2 \lambda_-^2}{\Delta} e^{2 \lambda_+ t} 
\left(1 - \frac{2\sqrt{\Delta}}{m\lambda_-} e^{-\lambda_+ t}\right)
\end{equation}
Hence 
\begin{equation}
G^2(t) \equiv  \frac{ \left(x_0 - X_0(t)\right)^2}{2\sigma^2} \approx
-\frac{m^2x_0^2 \lambda_-^2}{2\Delta} 
\frac{\Delta \lambda_+}{mk_B T \lambda_- }
\left(1 - \frac{2\sqrt{\Delta}}{m\lambda_-} e^{-\lambda_+ t}\right)
= \beta E \left(1 + \frac{4\sqrt{\Delta}}{\gamma + \sqrt{\Delta}} 
e^{-\lambda_+ t} \right)
\label{app:G2lim}
\end{equation}
where we have used $\lambda_+ \lambda_- = -K/m$ and $E=Kx_0^2/2$ for the 
barrier height.
From the previous equation we obtain the long time expansion of the 
derivative
\begin{equation}
G'(t) \approx  -\sqrt{\beta E} \frac{\sqrt{\Delta}}{\gamma + \sqrt{\Delta}} \lambda_+ 
e^{-\lambda_+ t}
\label{app:Gplim}
\end{equation}

\subsection{The overdamped regime}
\label{app:overdamped}

Having described the general underdamped case we illustrate now the
solution of the overdamped case. For the initial condition $x(0)=-x_0$
we find:
\begin{equation}
x(t) = -x_0 e^{\Omega t} + 
\frac{1}{\gamma} \int_0^t e^{\Omega (t-\tau)} \eta(\tau)
\end{equation}
where we have introduced the characteristic rate $\Omega \equiv
K/\gamma$. In the overdamped case there is a single rate. We also note
that $\lambda_+ \to \Omega$ in the overdamped case at strong friction
(see~\eqref{app:lambda_lim}).
The average of $x(t)$ corresponds to the deterministic trajectory
of a particle sliding down from the potential barrier
\begin{equation}
X(t) \equiv \left\langle x(t) \right\rangle = -x_0 e^{\Omega t}
\end{equation}
The variance is:
\begin{equation}
\sigma^2(t) \equiv 
\left\langle \left(x(t) - X(t) \right)^2 \right\rangle =  
\frac{k_BT}{K} \, \left( e^{2 \Omega t} - 1\right)
\end{equation}
from which we get:
\begin{equation}
G^2(t) \equiv \frac{\left(x_0 - X(t)\right)^2}{2 \sigma^2(t)}
= \frac{x_0^2 \left( e^{\Omega t} + 1\right)^2}
{\frac{2k_BT}{K} \left( e^{2 \Omega t} - 1\right)} = 
{\beta E} \, \frac{1+e^{-\Omega t}}{1 - e^{-\Omega t}}
\label{app:ovd}
\end{equation}
One can get to the same results by using the expressions obtained in the
overdamped case by formally taking the limit $m \to 0$.  In this limit
$\lambda_- t \to -\infty$ and $(\lambda_+ + \lambda_-) t \to -\infty$.
From Eq.~\eqref{app:sigma2} one gets:
\begin{equation}
\sigma^2 \to \frac{2m k_BT}{\Delta} 
\left( -\frac{\lambda_-}{2\lambda_+} e^{2 \lambda_+ t} +
\frac{\lambda_-}{2\lambda_+} +
\frac{\lambda_+}{2\lambda_-}  - 1
\right) =
-\frac{m  k_BT\lambda_-}{\Delta \lambda_+} 
\left[
e^{2 \lambda_+ t} - \left( \frac{\lambda_+ + \lambda_-}{\lambda_-}\right)^2
\right]
\to \frac{k_BT}{K} \left( e^{2 \Omega t} - 1\right)
\label{app:lim1}
\end{equation}
where we have used the fact that $\lambda_- \to -\infty$ and $m
\lambda_-/(\lambda_+ \Delta) \to 1/K$ in the limit $m\to 0$.
Using the same limiting behavior we find
\begin{equation}
\left( x_0 - X_0 (t) \right)^2 \to
\left( x_0 - \frac{m x_0 \lambda_-}{\sqrt{\Delta}} e^{\lambda_+ t}\right)^2
\to x_0^2 \left( 1 + e^{\Omega t} \right)^2
\label{app:lim2}
\end{equation}
Equations~\eqref{app:lim1} and \eqref{app:lim2} reproduce the overdamped
case, calculated directly in Eq.~\eqref{app:ovd}.

\section{Average TPT}
\setcounter{equation}{0}
\renewcommand{\theequation}{B\arabic{equation}}
\label{sec:app2}

We show how to derive equations (\ref{TPT_over}) and (\ref{TPT_under}).
The basic equation to be used is  (\ref{av_TPT}). 
\begin{itemize}
\item Overdamped case

The relation between $t$ and $G$ can be obtained by inverting relation (\ref{G2}). We have
\be
t=-\frac 1 \Omega \log \left ( \frac {G^2/\beta E -1}{ G^2/\beta E +1} \right)
\ee
In (\ref{av_TPT}), we make the change of variable
\be
x=G^2 - \beta E
\ee
and obtain
\be
\langle t_{TP} \rangle = -\frac{1}{\Omega} \frac{ \int_0^\infty \frac{dx}{\sqrt{1+ \frac{x}{\beta E}}} 
\left( \log x - \log \beta E - \log 2 - \log (1 + \frac{x}{2\beta E}) \right) e^{-x}}
{\int_0^\infty \frac{dx}{\sqrt{1+ \frac{x}{\beta E}}} e^{-x}}
\ee

Expanding this expression for large barrier $\beta E \gg 1$, we obtain
\be
\langle t_{TP} \rangle = \frac{\gamma}{K} 
\log \left(2 e^C \beta E \right) + O\left(\frac{1}{\beta E} \right)
\ee
where $C= - \int_0^{\infty} \log x \ e^{-x} \approx 0.577 215$ is the
Euler-Mascheroni constant.

\item Underdamped case

In that case, the relation between $G$ and $t$ cannot be inverted
analytically. However, we can use the asymptotic form of $G$ from
(\ref{app:G2lim})
\be
G=\sqrt \beta E \left( 1 + B e^{-\lambda_+t} \right)
\ee
where 
\be
B= \frac{2 \sqrt \Delta}{\gamma +  \sqrt \Delta}
\ee
We can now express $t$ as a function of $G$ as
\be
t = \frac{1}{\lambda_+} \left( \log B - \log \left( \frac{G}{\sqrt \beta E}-1 \right) \right)
\ee
which we insert in (\ref{av_TPT}). The calculations are very similar to
those of the overdamped case. Performing the change of variable $x=G^2 -
\beta E$ and expanding for high barrier $\beta E \gg 1$, we obtain the
asymptotic expansion
\be
\langle t_{TP} \rangle = \frac{1}{\lambda_+} 
\left( \log \beta E + \log 2 e^C B \right)+ O\left(\frac{1}{\beta E} \right)
\ee
where $C$ is defined above.  Note that this formula coincides with the
overdamped case in the limit when the mass vanishes.

\end{itemize}

\end{widetext}


%
\end{document}